\documentclass[english,reprint,aps,prl, superscriptaddress]{revtex4-1}
\usepackage[T1]{fontenc}
\usepackage[latin9]{inputenc}
\usepackage{amsmath}
\usepackage{amssymb}
\usepackage{graphicx}
\usepackage[final]{changes}

\makeatletter

\usepackage{babel}

\makeatother

\usepackage{babel}
\usepackage{tikz}
\usetikzlibrary{shapes,arrows}
\usetikzlibrary{arrows.meta}

\tikzstyle{decision} = [diamond, draw, fill=blue!20, 
text width=4.5em, text badly centered, node distance=3cm, inner sep=0pt]
\tikzstyle{block} = [rectangle, draw, fill=blue!30, 
text width=18.5em, text centered, rounded corners, minimum height=4em]
\tikzstyle{line} = [draw, -{Latex[length=0.3cm,width=0.5cm]}, ultra thick]
\tikzstyle{cloud} = [fill=red!50,draw, ellipse, node distance=3cm,
minimum height=2em]

\begin{document}

\title{Driven Imposters: Controlling Expectations in Many-Body Systems}
\date{\today}

\author{Gerard McCaul}
\email{gmccaul@tulane.edu}
\affiliation{Tulane University, New Orleans, LA 70118, USA}

\author{Christopher Orthodoxou}
\email{christopher.orthodoxou@kcl.ac.uk}
\affiliation{Department of Physics, King's College London, Strand, London, WC2R 2LS, U.K.}

\author{Kurt Jacobs} 
\affiliation{U.S. Army Research Laboratory, Computational and Information Sciences Directorate, Adelphi, Maryland 20783, USA} 
\affiliation{Department of Physics, University of Massachusetts at Boston, Boston, MA 02125, USA} 
\affiliation{Hearne Institute for Theoretical Physics, Louisiana State University, Baton Rouge, LA 70803, USA} 

\author{George H. Booth}
\email{george.booth@kcl.ac.uk}
\affiliation{Department of Physics, King's College London, Strand, London, WC2R 2LS, U.K.}

\author{Denys I. Bondar}
\email{dbondar@tulane.edu}
\affiliation{Tulane University, New Orleans, LA 70118, USA}

\begin{abstract}
We present a framework for controlling the observables of a general correlated electron system driven by an incident laser field. The approach provides a prescription for the driving required to generate an arbitrary predetermined evolution for the expectation value of a chosen observable, together with a constraint on the maximum size of this expectation.
To demonstrate this, we determine the laser fields required to exactly control the current in a Fermi-Hubbard system under a range of model parameters, fully controlling the non-linear high-harmonic generation and optically observed electron dynamics in the system. This is achieved for both the uncorrelated metallic-like state and deep in the strongly-correlated Mott insulating regime, flipping the optical responses of the two systems so as to mimic the other, creating `driven imposters'. We also present a general framework for the control of other dynamical variables, opening a new route for the design of driven materials with customized properties. 
\end{abstract}
\maketitle

\paragraph{Introduction:-} Ohm's law is one of the most ubiquitous relationships in all of physics, beginning as an empirical law \citep{Ohm} before both the Drude and free electron models provided a quantitative justification for its existence\citep{neilashcroft1976}. In the Ohmic regime the relationship between a driving field and the observed current is linear, so that for a given current there is a unique and trivial solution for the control field required to produce it. While it is possible to find examples of Ohm's law persisting down to the atomic scale \citep{Weber64}, physical systems abound with phenomena such as persistent currents \citep{Bleszynski-Jayich272} and High Harmonic Generation (HHG)  \citep{Ghimire2012, Ghimire2011a, Murakami2018} where the linear relationship breaks down. This has important consequences for the control of such systems, where the manipulation of an expectation with a non-linear dependence on a control field presents both significant challenges and opportunities for exploitation \citep{HHGcontrol,RabitzScience}. A diverse array of strategies have been previously proposed to address this, including both optimal \citep{Werschnik2007, Serban2005,Kosloffoptimalconstraints,PhysRevLett.106.190501} and local \citep{Kosloff1992,Koslofflocalcontrol} quantum control. 

The ability to manipulate expectation values in this way is highly desirable, with obvious benefits. It presents an opportunity in both materials science and chemistry to substitute simpler and cheaper compounds that can mimic the desired properties of more expensive materials\citep{Zhang2005,doi:10.1021/ja00081a012,doi:10.1021/ar00028a010,C4RA01210K,doi:10.1021/am506611j}. A concrete example of the need for control strategies beyond the linear regime can be found in recent experimental \citep{Nicoletti2018} and theoretical \citep{Schlawin2019} work which demonstrates photo-induced superconductivity in materials above their critical temperature $T_c$ \citep{Cavalleri}. This raises the possibility of designing laser pulses that induce superconductivity or other dynamical phase transitions,
but to do so requires the ability to control expectations beyond the linear regime.

In this letter, we present a method for time-dependent control of expectations within correlated many-body electronic systems, when systems observables have a highly non-linear dependence on the control field $E(t)$. Since this method allows an expectation value to follow (or `track') an essentially arbitrary function of time, up to a scaling factor, we will refer to it as \textit{tracking control} \citep{PhysRevA.72.023416,PhysRevA.98.043429,PhysRevA.84.022326,Campos2017, doi:10.1063/1.1582847,doi:10.1063/1.477857} \added{(for further details on this and other control strategies, see Ref.~\onlinecite{companionlong})}.  

\added{One of the principal advantages of tracking control is its computational efficiency as compared to the iterative optimisation of optimal control \citep{PhysRevA.98.043429}. Exact tracking control can however suffer from singularities in the control field as a consequence of specifying a track inconsistent with physical dynamics. The model presented here possesses several key advantages to address this. By working in a finite dimensional context, this method is 
explicitly applicable to many-electron solid-state systems on a discrete lattice. The tracking equations derived from this model are insensitive as to whether the system is evolved as a closed or open system, and it is also possible to determine the precise constraints necessary to avoid singularities and guarantee a unique evolution of the system. This is particularly desirable, as it removes one of the main obstacles to tracking control - the ability to determine if a trajectory is physically realisable}. 
We test the new method in the highly non-linear regime of HHG in the Hubbard model, using it both to induce arbitrarily designed currents, as well as creating `driven imposters' where the optical spectrum of one material mimics that of another. 

\paragraph{Summary of Results:- \label{sec:Model}} 
Our goal is to implement a tracking control \citep{doi:10.1063/1.1582847} model for a general $N$-electron system subjected to a laser pulse.
Specifically, we wish to calculate a control field 
such that the trajectory of an expectation $\left<\hat{O}(t)\right>$ under the Hamiltonian is described by some desired function $O_T(t)$ \citep{PhysRevA.72.023416,PhysRevA.98.043429,PhysRevA.84.022326,Campos2017}. 
While the general tracking strategy detailed here is suitable for any Hamiltonian (see Ref.~\onlinecite{companionlong}), 
we will focus on the discrete 1D Fermi-Hubbard model as a paradigmatic model of strongly correlated electron systems, given by 
\begin{align}
\hat{H}\text{\ensuremath{\left(t\right)}}= & -t_{0}\sum_{j\sigma}\text{\ensuremath{\left({\rm e}^{-i\Phi\left(t\right)}\hat{c}_{j\sigma}^{\dagger}\hat{c}_{j+1\sigma}+{\rm e}^{i\Phi\left(t\right)}\hat{c}_{j+1\sigma}^{\dagger}\hat{c}_{j\sigma}\right)}}\nonumber \\
 & +U\sum_{j}\hat{c}_{j\uparrow}^{\dagger}\hat{c}_{j\uparrow}\hat{c}_{j\downarrow}^{\dagger}\hat{c}_{j\downarrow},
\label{eq:Hamiltonian}\end{align}
where the correlated physics is induced by the on-site repulsive $U$ term \citep{Tasaki1998}, with the phase $\Phi\text{\ensuremath{\left(t\right)}}=aA(t)$, describing the applied field, where $a$ is the lattice constant and $A(t)$ is the field vector potential.

Our aim is to have the expectation of the current operator $\hat{J}(t)$ \citep{fabianessler2005} ,
\begin{equation}
\hat{J}(t)=-iat_{0}\sum_{j\sigma}\left({\rm e}^{-i\Phi\left(t\right)}\hat{c}_{j\sigma}^{\dagger}\hat{c}_{j+1\sigma}-{\rm H.c.}\right)\label{eq:currentoperator},
\end{equation}
track some predetermined target function $J_T (t)$, such that $\langle \hat{J}(t)\rangle =J_T (t)$. Imposing this constraint on the system evolution $i\frac{{\rm d}\left|\psi\right>}{{\rm d}t}=\hat{H}\left(t\right)\left|\psi\right>$ is equivalent to evolving the wave function via a  non-linear evolution given by
\begin{align}
    i\frac{{\rm d}\left|\psi\right>}{{\rm d}t}=\hat{H}_T\left(J_T(t), \psi\right)\left|\psi\right>, \label{eq:eqnofmotion}
\end{align}
where $\hat{H}_T (J_T(t),\psi)$ is the `tracking Hamiltonian' which takes the target function $J_T(t)$ as a parameter, and acquires a dependence on the current state of the system, $\psi$. An explicit form for $\hat{H}_T (J_T(t),\psi)$ can then be found as,
\begin{align}
\hat{H}_{T}\left(J_T(t), \psi\right)  =&\sum_{\sigma,j}P_+{\rm e}^{-i\theta\left(\psi \right)}\hat{c}_{j\sigma}^{\dagger}\hat{c}_{j+1\sigma},  \nonumber \\
+&\sum_{\sigma,j}P_-{\rm e}^{i\theta\left(\psi\right)}\hat{c}_{j+1\sigma}^{\dagger}\hat{c}_{j\sigma}+U\sum_{j}\hat{c}_{j\uparrow}^{\dagger}\hat{c}_{j\uparrow}\hat{c}_{j\downarrow}^{\dagger}\hat{c}_{j\downarrow}, \label{eq:trackingHamiltonian1} \\
P_{\pm} =&-t_{0}\left(\sqrt{1-X^2(t,\psi)}\pm iX(t,\psi)\right), \label{eq:trackingHamiltonian2} \\
X(t,\psi)=&\frac{J_T\left(t\right)}{2at_{0}R\left(\psi\right)}.
\label{eq:trackingHamiltonian3}
\end{align}
The $\psi$ dependence is defined by the neighbour hopping expectation in a polar form,
\begin{equation}
\left\langle \psi \left|\sum_{j\sigma} \hat{c}_{j\sigma}^{\dagger}\hat{c}_{j+1\sigma}\right| \psi \right\rangle =R\left(\psi\right){\rm e}^{i\theta\left(\psi\right)}. \label{neighbourexpectation}
\end{equation}
Under this tracking Hamiltonian, the evolution of $\psi$ is equivalent to that given by the Hamiltonian in Eq.~(\ref{eq:Hamiltonian}) under the action of a field,
\begin{align}
\Phi(t)=\Phi_T(t)=\arcsin\left(\frac{-J_T(t)}{2at_{0}R(\psi)}\right)+\theta(\psi) \label{eq:tracking_phi}.
\end{align} 
The form of this tracking Hamiltonian imposes some constraints on the currents that can be tracked successfully. In order to ensure that the evolution is unitary, and that Eq.~(\ref{eq:eqnofmotion}) has a unique solution for $\psi$, it is sufficient to require $\forall ~\psi$, and that
\begin{align}
\added{\left|X(t,\psi)\right|} &\added{<1-\epsilon_1} \label{eq:TrackingCondition} \\
\added{R(\psi)} &\added{>\epsilon_2 ,} \label{eq:TC2}
\end{align}
where $\epsilon_{1/2}$ are small finite constants. When these constraints are satisfied, the tracking Hamiltonian is not only Hermitian (i.e., $P^\dagger_+=P_-$), but guarantees a unique solution for $\psi$ \citep{kentnagle2011} despite the non-linear character of its evolution. A full derivation of this result along with a physical interpretation of the constraints is given in Ref.~\onlinecite{companionlong}. This result, derived from functional analysis, stands in sharp contrast to some discrete models, in which multiple solutions for tracking are possible \citep{Rabitzmultiple}.

\added{Importantly, the constraints above define the limits on the size of expectations it is possible to produce with physically realisable control fields.} While in principle the constraint of Eq.~(\ref{eq:TrackingCondition}) is a highly non-linear inequality in $\psi$, in practice it is relatively easy to satisfy via a heuristic scaling of the target to be tracked, as these constraints limit only the peak amplitude of current in the evolution and otherwise allow for any function to be tracked when appropriately scaled. If one is concerned only with reproducing the shape of the target current, then using a scaled target $J_s(t)=kJ_T(t)$ such that $\left|J_s\left(t\right)\right|<2at_{0}R\left(\psi\right)$  will allow tracking without problem. Alternately, if one treats the lattice constant $a$ as a tunable parameter, this can always be set for the tracking system so as to satisfy Eq.~(\ref{eq:TrackingCondition}). This approach also ensures the avoidance of singularities in the trajectories, which have often afflicted other tracking control approaches\citep{doi:10.1063/1.1582847, doi:10.1137/0325030, PhysRevA.98.043429}.


This tracking strategy can be generalized for an arbitrary expectation value of interest. Tracking of an
observable $\hat{O}$ such that $\langle \hat{O}\rangle=O_T(t)$, one requires the following expectations
	\begin{align}
	    R_{O}(\psi){\rm e}^{i\theta_{O}(\psi)}=&\sum_{\sigma,j}\left\langle \left[\hat{c}_{j\sigma}^{\dagger}\hat{c}_{j+1\sigma},\hat{\hat{O}}\right]\right\rangle, \\
	    B(\psi)=&-iU\sum_{j}\left\langle \left[\hat{c}_{j\uparrow}^{\dagger}\hat{c}_{j\uparrow}\hat{c}_{j\downarrow}^{\dagger}\hat{c}_{j\downarrow},\hat{O}\right]\right\rangle.
	\end{align} 
Using these expectations, Eqs.~(\ref{eq:trackingHamiltonian1})-(\ref{eq:trackingHamiltonian2}) may be used to track the observable using the substitutions $J_T(t)= O_T(t)$, $ R(\psi){\rm e}^{i\theta(\psi)}=  R_{O}(\psi){\rm e}^{i\theta_{O}(\psi)}$, with 
\begin{equation}
X(t,\psi) = \frac{\frac{{\rm d} O_T(t)}{{\rm d}t}-B(\psi)}{2t_{0}R_{O}(\psi)},
\end{equation}
and constraints on $X(t,\psi)$ given by Eq.~(\ref{eq:TrackingCondition}) in the same fashion as current tracking. More details deriving tracking of arbitrary observables is given in Ref.~\onlinecite{companionlong}. 

\added{Finally, we note that the expression for the tracking field in Eq.\eqref{eq:tracking_phi} depends only implicitly on the system evolution through $R(\psi)$ and $\theta(\psi)$, and is derived only through the definition of $\hat{J}(t)$. This means that the definition of the tracking field (and therefore the tracking Hamiltonian) is insensitive to whether the system is evolving in a Liouvillian (closed) or Lindbladian (open) manner. } 

\paragraph{Reference systems:-} In order to test our tracking strategy we consider the 1D Fermi-Hubbard model at both $U/t_0=0$, where the system is in a metallic Tomonaga-Luttinger liquid phase \citep{PhysRevLett.45.1358}, and $U/t_0=7$ where the system is deep in a Mott insulating regime with large optical bandgap \citep{PhysRevLett.85.3910}. We consider an $L=10$ site Hubbard chain with periodic boundary conditions with an average of one electron per site, a hopping parameter of $t_0=0.52$~eV, and lattice constant of $a=4$~\AA. While this system size is not at the thermodynamic limit of the model, it is sufficient to allow for demonstration of the method and qualitative agreement with the bulk limit \citep{Silva2018}, whilst allowing for an exact propagation of the wave function so as not to introduce errors from an approximate time-evolution. To each reference system we apply a laser pulse of duration $N=10$ periods, described by the Peierls phase
\begin{align}
    \Phi(t)=a\frac{E_0}{\omega_0}\sin^2\left( \frac{\omega_0 t}{2N}\right)\sin(\omega_0t).
    \label{eq:refphase}
\end{align}
This is related to the electric field $E(t)$ via $aE(t)=-\frac{{\rm d} \Phi}{{\rm d}t}$. The pulse parameters are chosen as experimentally feasible field amplitudes of $E_0=10$~MV/cm with frequency $\omega_0=32.9$~THz \citep{Hohenleutner2015}. 

Driving with this field produces a highly non-linear response of these reference systems. A particular manifestation of this is the phenomenon of high-harmonic generation (HHG) \citep{Ghimire2012, Ghimire2011a, Murakami2018}, where the incident laser field produces high-order harmonics in the current, and drastically alters its electronic properties \citep{Ghimire2011}. This phenomenon has proven to be a useful tool, enabling molecular orbital tomography \citep{Pepin2004} and femtosecond resolution imaging of strongly correlated systems \citep{Silva2018}. HHG potentially even offers a route to studying dynamics in the attosecond regime \citep{RevModPhys.81.163}, as well as precise chiral spectroscopy \citep{PhysRevX.9.031002}.

While the undriven 1D Hubbard model is an insulator for all $U>0$ \citep{PhysRevLett.20.1445}, when a laser pulse is applied, there are two distinct phases. The driven system may be in a Mott insulating phase, or -- if the incident field is of sufficient amplitude -- it undergoes a dielectric breakdown and becomes conducting \citep{floriangebhard2010}. In the regime of $\omega_0 <U$ where the linear response of the system cannot excite electrons across the gap, the dominant mechanism for this breakdown is non-linear quantum tunnelling, with an associated critical field amplitude \citep{PhysRevB.86.075148} 
\begin{equation}
    E_{\rm{th}}\sim\frac{\Delta}{2\xi}, \label{eq:threshold}
\end{equation}
where $\Delta$ is the Mott gap \citep{PhysRevB.79.245120}, and $\xi$ is the doublon-hole correlation length \citep{PhysRevB.86.075148}. This threshold can be rationalized as the field strength required to separate a charge pair by a large enough distance to distinguish them.

\begin{figure}
\begin{center}
\includegraphics[width=1.1\columnwidth]{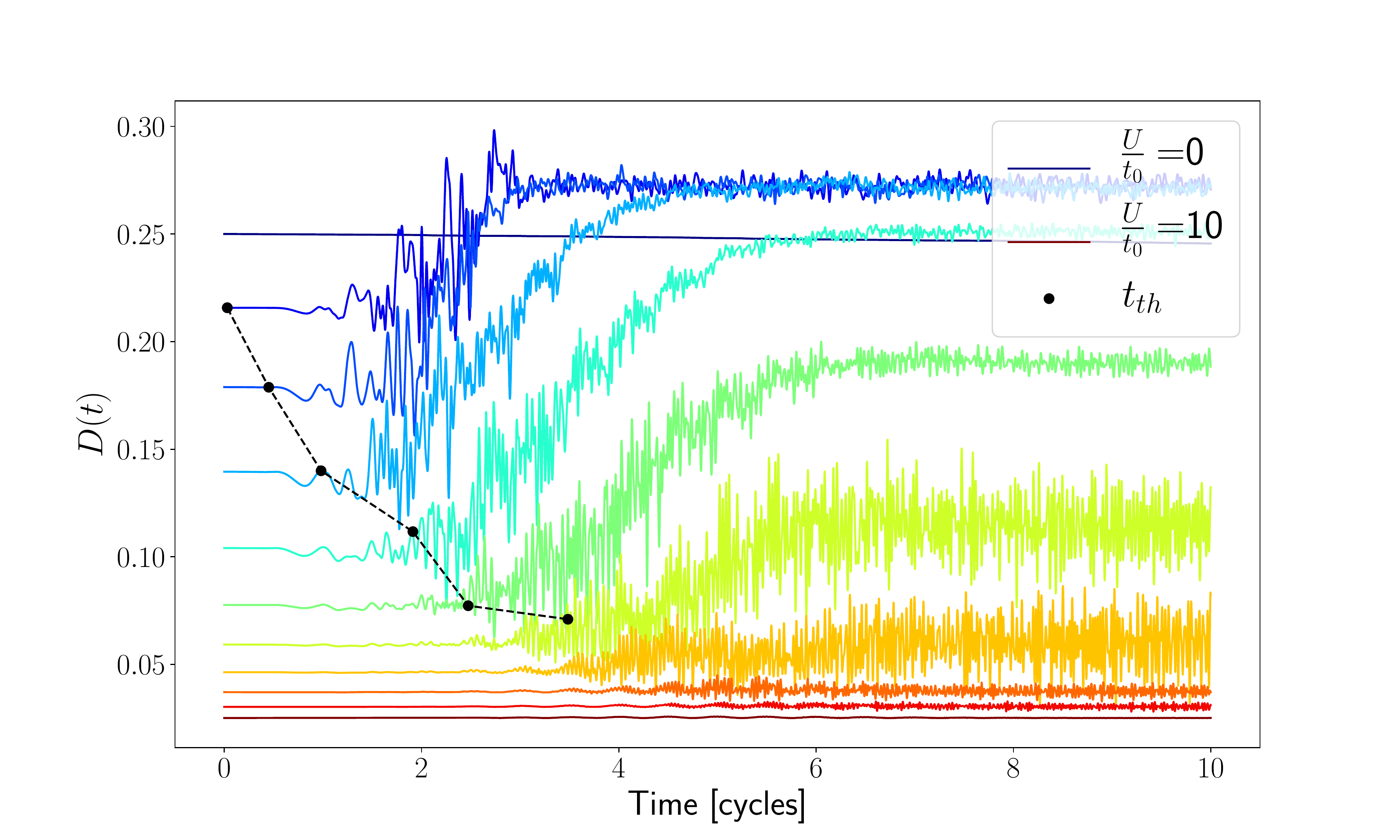}\end{center}\caption{Doublon occupation for increasing correlation strengths of $U/t_0$ under the action of the laser. Simulations are  distinguished by unit increments in $U/t_0$, from $U=0$ to $U=10t_0$. Black points indicate the time when dielectric breakdown is predicted to occur, and $U=7t_0$ is the first plotted value for which the laser field has an insufficient amplitude for causing a breakdown.} 
\label{fig:doublonref}
\end{figure}

In the conducting system, the charge-carriers are doublons and holes \citep{fabianessler2005}, characterized by the \emph{doublon occupation} \citep{PhysRevB.86.075148} of,
\begin{align}
    D(t)=\frac{1}{L}\left<\sum_{j}\hat{c}_{j\uparrow}^{\dagger}\hat{c}_{j\uparrow}\hat{c}_{j\downarrow}^{\dagger}\hat{c}_{j\downarrow} \right>.
\end{align}
If the breakdown threshold is reached, the density of these charge carriers increases. 
Fig. \ref{fig:doublonref} shows that the numerical rise in the doublon occupation occurs at roughly time $t_{th}$, estimated to be when the incident field first meets the critical threshold, assuming that $\Delta(U)$ is not too large for this to be reached. This threshold time is given by the solution to
\begin{align}
\frac{\Delta}{2 E_0 \xi}=& \sin^2\left( \frac{\omega_0 t_{th}}{2N} \right) \cos(\omega_0t_{th}) \nonumber \\ &+\frac{1}{2N}\sin \left( \frac{\omega_0 t_{th}}{N} \right) \sin(\omega_0t_{th}), \label{eq:t_th}
\end{align}
where analytic expressions in the thermodynamic limit are used for both $\Delta$ \citep{LIEB20031, floriangebhard2010} and $\xi$ \citep{PhysRevB.86.075148,PhysRevB.48.1409}, and exists provided $U\lesssim 6.9t_0$. Beyond this threshold the reference driving field's amplitude is insufficient to cause a dielectric breakdown. This breakdown behaviour and its effect on $D(t)$ can be seen in Fig. \ref{fig:doublonref}. 

\begin{figure}
\begin{center}
\includegraphics[width=1\columnwidth]{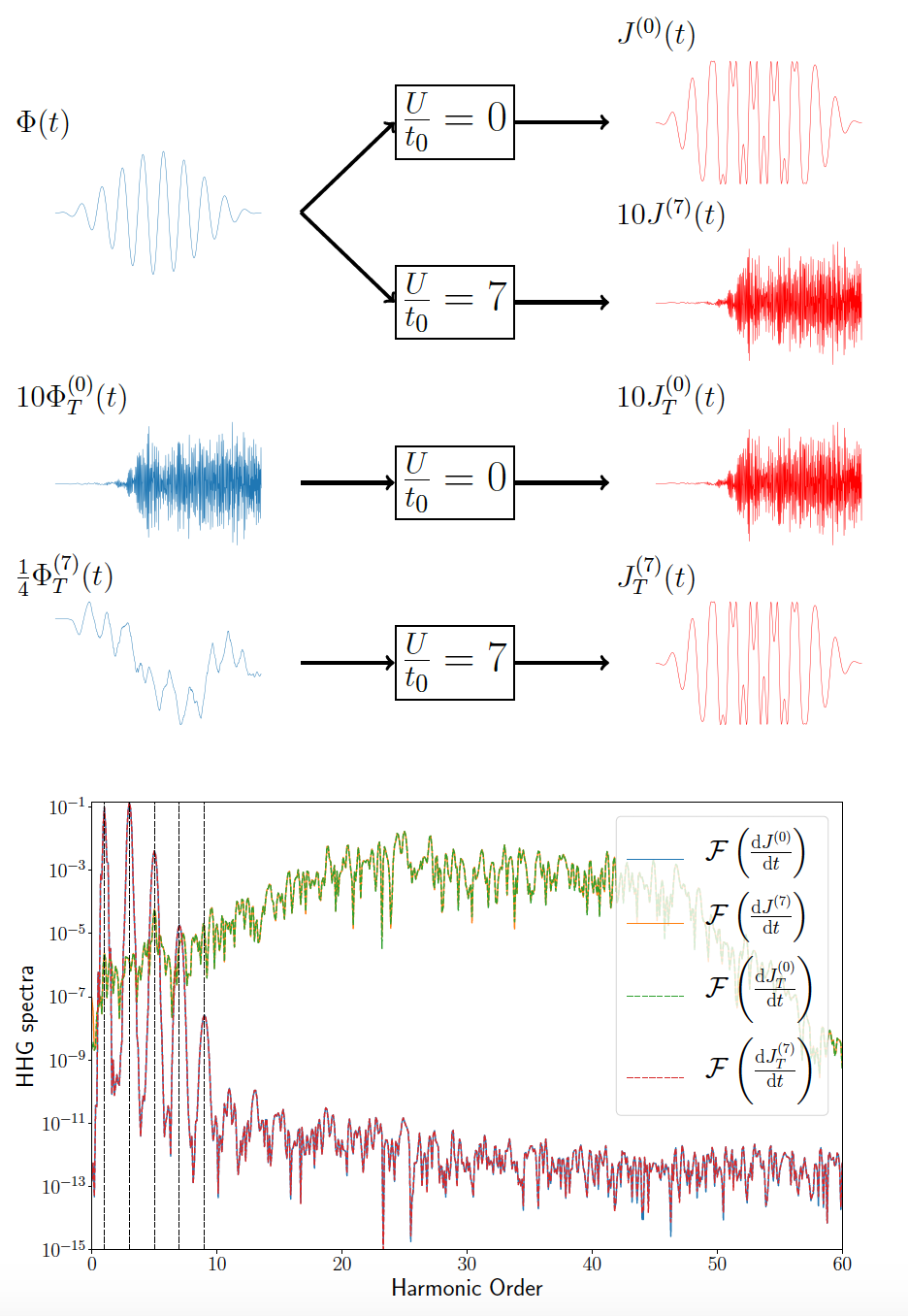}\end{center}\caption{Using tracking, it is possible to make the HHG spectra of one system mimic the other. Here tracking has been implemented to swap the optical characteristics of two systems, i.e. $J_T^{(0)}(t)=J^{(7)}(t)$ and $J_T^{(7)}(t)=J^{(0)}(t)$. The top section shows the original and tracked control fields and currents in the time domain, while the bottom section demonstrates the strategy's success in mimicking spectra.}
\label{fig:tracking_spectra}
\end{figure}

One consequence of the driving field's creation of charge excitations is in the optical response, where higher harmonics are generated in the HHG spectrum of the dipole acceleration ($\frac{{\rm d}J}{{\rm d}t}$). The two contrasting regimes are shown in Fig. \ref{fig:tracking_spectra}. For $U=0$ the system is a conductor and exhibits well-defined peaks at odd harmonics, as observed in other mono-band tight-binding models \citep{Schubert2014}. In contrast, at $U=7t_0$ the Mott gap is such that $E_{\rm th}(U) > E_0$ and the system is unable to create charge carriers even under driving. In the HHG spectrum, the low-order harmonics are suppressed and effective intra-band high-harmonic generation dominates \citep{Hawkins2013}, broadening the spectrum, with a peak at $N \sim U / \omega_0$  \citep{Silva2018}. 

\paragraph{Material Mimicry:-}
A key target application for tracking control is the ability to make one material mimic the spectral behaviour of another. To demonstrate this, we use the tracking strategy to make the $U=0$ system mimic the HHG spectrum of the $U=7t_0$ system and vice versa. 
The observed current will be labeled with a superscript to indicate the $U/t_0$ value used, e.g., the current expectation for the $U=0$ model is labeled $J^{(0)}(t)$, while for $U=7t_0$ the current expectation is $J^{(7)}(t)$. Finally, we will label the expectations generated in the presence of the tracking field with a subscript $T$. For example, the current expectation of the $U=0$ system with tracking used to imitate the $U=7t_0$ system is $J_T^{(0)}(t)=J^{(7)}(t)$.

An important caveat here is that directly reproducing the conducting system's current in the insulating system is complicated by the fact that the maximum current a system may generate is proportional to $R(t)$, which will in general be much greater in the conducting system. Trying to track $J^{(0)}(t)$ in the insulating system directly violates the tracking condition given by Eq.~(\ref{eq:TrackingCondition}). To remedy this, the lattice constant in the tracked system is scaled to a value $a_T^{(7)}=60a^{(0)}$, such that Eq.~(\ref{eq:TrackingCondition}) is obeyed at all times. Alternatively, one could simply scale $J^{(0)}(t)$ for tracking, while still retaining the essential spectral features of the conducting limit, i.e. tightly focused peaks around odd integer overtones of the driving frequency.

\begin{figure}
\includegraphics[width=1.1\columnwidth]{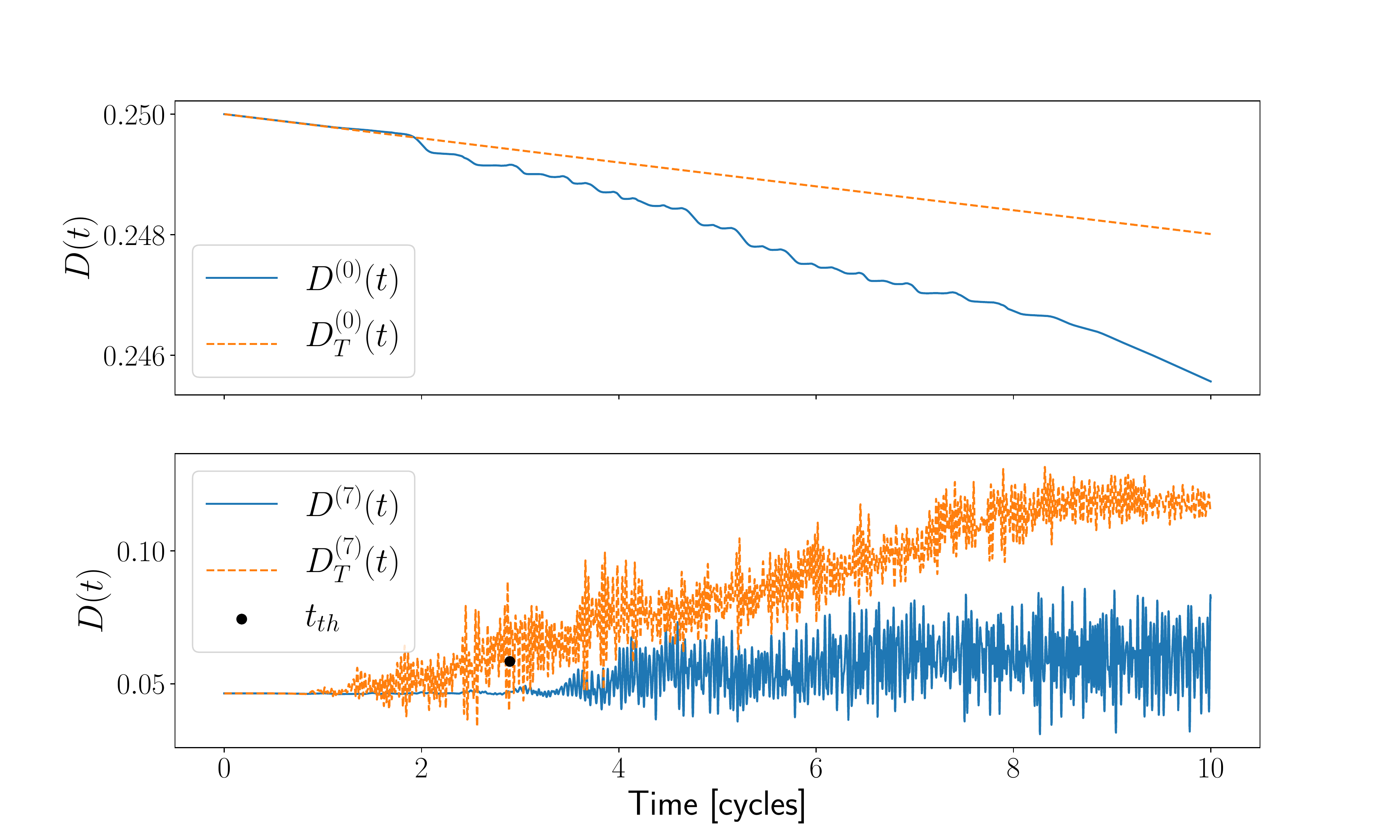}\caption{The doublon density of the $U=0$ (upper plot) and $U=7t_0$ (lower plot) systems under the reference HHG field $D(t)$, or the tracking field to mimic the current of the other system $D_T(t)$. In the $U=0$ model, the doublon density is largely insensitive to the tracking field and resulting change in current, i.e. $D_T^{(0)}(t) \approx D^{(0)}(t)$. Conversely, at high $U/t_0$, the peak amplitude of the tracking control field needed to mimic the spectrum of the $U=0$ conducting system is large enough to cause a dielectric breakdown. This breakdown time for $D_T^{(7)}(t)$ is calculated via Eq.~(\ref{eq:threshold}), and demonstrates that tracking the spectrum of a conductor from the Mott state necessitates a breakdown of the insulating state as measured by the doublon occupation.}
\label{fig:doublon_tracking}
\end{figure}

Fig. \ref{fig:tracking_spectra} shows the success of the tracking strategy in spectral mimicry, where each material's reference HHG spectra can be tracked in the other. 
While current tracking is used to make the one material imitate the dipole acceleration spectrum of another, the doublon occupation (shown in Fig. \ref{fig:doublon_tracking}) is not explicitly tracked here, and provides an alternate characterization of the system state. This reveals that even while imitating the $U=7t_0$ current, the doublon occupation in the $U=0$ system indicates that it remains in the conducting limit, where $D^{(0)} \sim D_T^{(0)}$. This is to be expected, as running a small current through a conducting system would not change its conductive property. However, in the Mott insulating $U=7t_0$ system, a more dramatic change has occurred between the reference and tracked systems in order to mimic the spectrum of the $U=0$ conducting system. The tracking system must exceed the dielectric breakdown threshold given by Eq.~(\ref{eq:threshold}) in order to ensure enough mobile charge carriers to generate sufficient current.
The result of this is that $D_T^{(7)}$ exhibits a rise in doublon density characteristic of this dielectric breakdown. Importantly, the same qualitative behaviour also occurs when one instead chooses to scale the target current $J_s(t)=kJ_T(t)$ rather than the lattice constant $a$. This breakdown is confirmed by calculation of the time at which $\Phi^{(7)}_T(t)$ exceeds the threshold associated with $U=7t_0$ and is also shown in Fig. \ref{fig:doublon_tracking}.  

\begin{figure}
\begin{center}
\includegraphics[width=1\columnwidth]{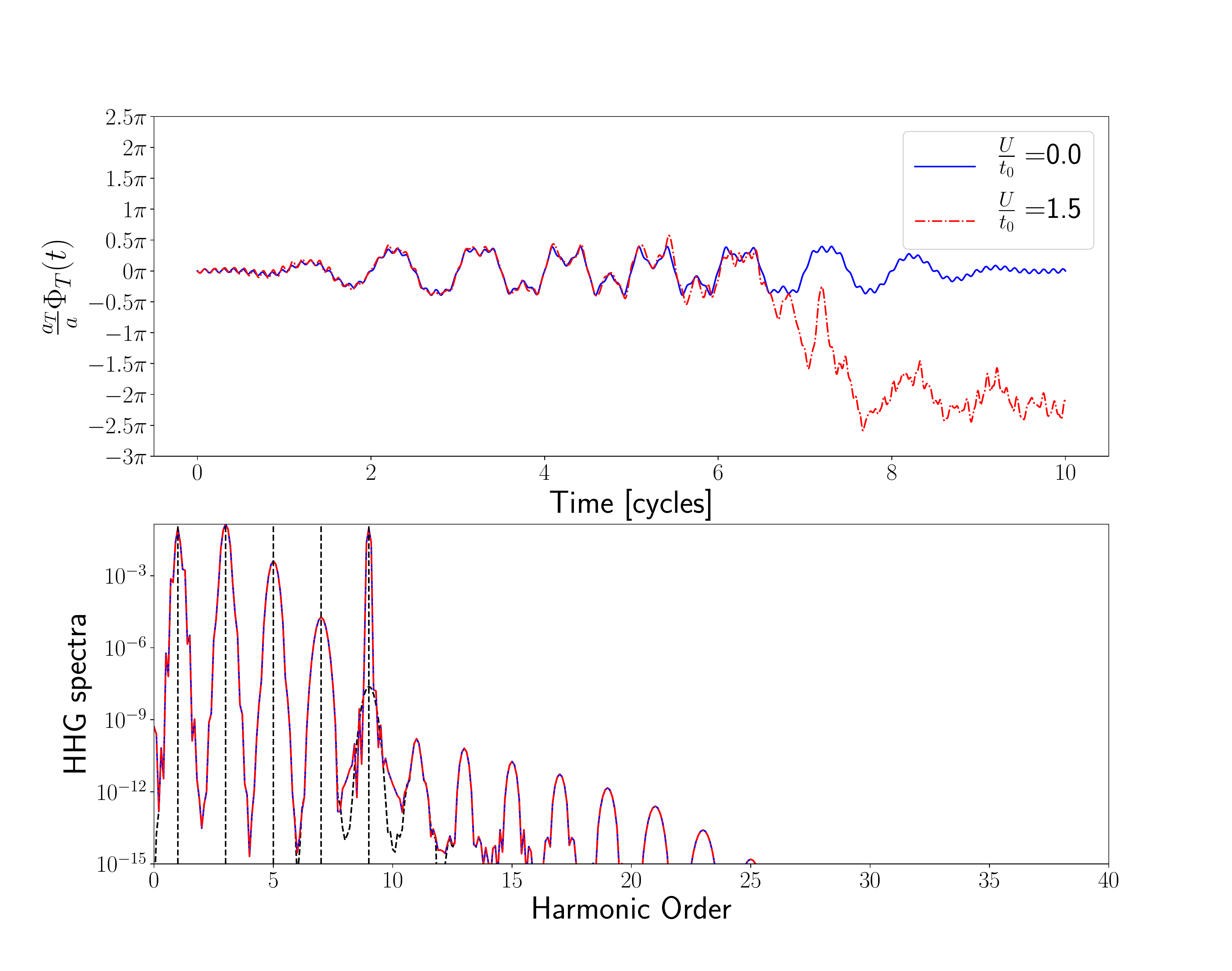}\end{center}\caption{\added{Using tracking, it is possible to boost the yield of a higher  (ninth) harmonic by tracking a synthetic spectrum. The upper panel shows the necessary control-field needed to reproduce this current is sensitive to the correlation strength.}}
\label{fig:switchtracking}
\end{figure}

\paragraph{Enhancing harmonics with arbitrary control:-}
\added{With the arbitrary control provided by tracking, it is possible to address a longstanding goal for the manipulation of systems exhibiting HHG. Namely, enhancing the yield of a specific high harmonic \citep{HHGenhance1,HHGenhance2,HHGenhance3}.
In Fig.~\ref{fig:switchtracking} we show the result of applying the tracking algorithm to generate a current that matches a synthetic spectrum where the ninth harmonic in the spectrum has been boosted to a level comparable with the first harmonic. The tracking phase $\Phi_T(t)$ necessary to produce this boosted yield is also shown at several interaction strengths.}

\paragraph{Discussion:-}
We have demonstrated a strategy for arbitrarily manipulating the current, (and therefore HHG spectra) of a strongly-correlated system. Several applications of this technique were discussed. Tracking control on many-electron systems provides a route to exerting fine control over the HHG spectrum of a strongly-correlated system.  Previous experiments have been able to effectively characterise both a THz control field and the optical spectrum it induces \citep{Sommer2016}, and the experimental feasibility of the scheme presented here is discussed in Ref.~\onlinecite{companionlong}. \added{We find that it is possible to produce a reasonable approximation of the control fields using only two additional frequencies, which in turn capture the essential qualitative behaviour of the tracked currents. While this is encouraging, it is important to remain aware of the difficulties of long-term control over transient phenomena, as decoherence and errors in the initial setup accumulate larger effects over time. Neverthless, } given the utility of HHG for the resolution of ultrafast many-body dynamics \citep{Silva2018}, we believe the approach presented here provides a potential route to controlling system dynamics on a sub-femtosecond time-scale.

\begin{acknowledgments}
\paragraph{Acknowledgments:-}
    G.M. and D.I.B. are supported by Air Force Office of Scientific Research (AFOSR) Young Investigator Research Program (grant FA9550-16-1-0254) and the Army Research Office (ARO) (grant W911NF-19-1-0377). The views and conclusions contained in this document are those of the authors and should not be interpreted as representing the official policies, either expressed or implied, of AFOSR, ARO, or the U.S. Government. The U.S. Government is authorized to reproduce and distribute reprints for Government  purposes notwithstanding any copyright notation herein.
    
    G.H.B and C.O. acknowledge funding by the Engineering and Physical Sciences Research Council (EPSRC) through the Centre for Doctoral Training ``Cross Disciplinary Approaches to Non-Equilibrium Systems" (CANES, Grant No. EP/L015854/1). G.H.B. gratefully acknowledges support from the Royal Society via a University Research Fellowship, and funding from the Air Force Office of Scientific Research via grant number FA9550-18-1-0515. The project has received funding from the European Union's Horizon 2020 research and innovation programme under grant agreement No. 759063.
\end{acknowledgments}


\begin{thebibliography}{53}%
	\makeatletter
	\providecommand \@ifxundefined [1]{%
		\@ifx{#1\undefined}
	}%
	\providecommand \@ifnum [1]{%
		\ifnum #1\expandafter \@firstoftwo
		\else \expandafter \@secondoftwo
		\fi
	}%
	\providecommand \@ifx [1]{%
		\ifx #1\expandafter \@firstoftwo
		\else \expandafter \@secondoftwo
		\fi
	}%
	\providecommand \natexlab [1]{#1}%
	\providecommand \enquote  [1]{``#1''}%
	\providecommand \bibnamefont  [1]{#1}%
	\providecommand \bibfnamefont [1]{#1}%
	\providecommand \citenamefont [1]{#1}%
	\providecommand \href@noop [0]{\@secondoftwo}%
	\providecommand \href [0]{\begingroup \@sanitize@url \@href}%
	\providecommand \@href[1]{\@@startlink{#1}\@@href}%
	\providecommand \@@href[1]{\endgroup#1\@@endlink}%
	\providecommand \@sanitize@url [0]{\catcode `\\12\catcode `\$12\catcode
		`\&12\catcode `\#12\catcode `\^12\catcode `\_12\catcode `\%12\relax}%
	\providecommand \@@startlink[1]{}%
	\providecommand \@@endlink[0]{}%
	\providecommand \url  [0]{\begingroup\@sanitize@url \@url }%
	\providecommand \@url [1]{\endgroup\@href {#1}{\urlprefix }}%
	\providecommand \urlprefix  [0]{URL }%
	\providecommand \Eprint [0]{\href }%
	\providecommand \doibase [0]{http://dx.doi.org/}%
	\providecommand \selectlanguage [0]{\@gobble}%
	\providecommand \bibinfo  [0]{\@secondoftwo}%
	\providecommand \bibfield  [0]{\@secondoftwo}%
	\providecommand \translation [1]{[#1]}%
	\providecommand \BibitemOpen [0]{}%
	\providecommand \bibitemStop [0]{}%
	\providecommand \bibitemNoStop [0]{.\EOS\space}%
	\providecommand \EOS [0]{\spacefactor3000\relax}%
	\providecommand \BibitemShut  [1]{\csname bibitem#1\endcsname}%
	\let\auto@bib@innerbib\@empty
	\bibitem [{\citenamefont {Ohm}(1827)}]{Ohm}%
	\BibitemOpen
	\bibfield  {author} {\bibinfo {author} {\bibfnamefont {G.}~\bibnamefont
			{Ohm}},\ }\href@noop {} {\emph {\bibinfo {title} {Die Galvanische Kette,
				Mathematisch Bearbeitet}}}\ (\bibinfo  {publisher} {Riemann},\ \bibinfo
	{year} {1827})\BibitemShut {NoStop}%
	\bibitem [{\citenamefont {Ashcroft}(1976)}]{neilashcroft1976}%
	\BibitemOpen
	\bibfield  {author} {\bibinfo {author} {\bibfnamefont {N.~W.}\ \bibnamefont
			{Ashcroft}},\ }\href {https://www.xarg.org/ref/a/0030839939/} {\emph
		{\bibinfo {title} {Solid State Physics}}}\ (\bibinfo  {publisher} {Cengage
		Learning},\ \bibinfo {year} {1976})\BibitemShut {NoStop}%
	\bibitem [{\citenamefont {Weber}\ \emph {et~al.}(2012)\citenamefont {Weber},
		\citenamefont {Mahapatra}, \citenamefont {Ryu}, \citenamefont {Lee},
		\citenamefont {Fuhrer}, \citenamefont {Reusch}, \citenamefont {Thompson},
		\citenamefont {Lee}, \citenamefont {Klimeck}, \citenamefont {Hollenberg},\
		and\ \citenamefont {Simmons}}]{Weber64}%
	\BibitemOpen
	\bibfield  {author} {\bibinfo {author} {\bibfnamefont {B.}~\bibnamefont
			{Weber}}, \bibinfo {author} {\bibfnamefont {S.}~\bibnamefont {Mahapatra}},
		\bibinfo {author} {\bibfnamefont {H.}~\bibnamefont {Ryu}}, \bibinfo {author}
		{\bibfnamefont {S.}~\bibnamefont {Lee}}, \bibinfo {author} {\bibfnamefont
			{A.}~\bibnamefont {Fuhrer}}, \bibinfo {author} {\bibfnamefont {T.~C.~G.}\
			\bibnamefont {Reusch}}, \bibinfo {author} {\bibfnamefont {D.~L.}\
			\bibnamefont {Thompson}}, \bibinfo {author} {\bibfnamefont {W.~C.~T.}\
			\bibnamefont {Lee}}, \bibinfo {author} {\bibfnamefont {G.}~\bibnamefont
			{Klimeck}}, \bibinfo {author} {\bibfnamefont {L.~C.~L.}\ \bibnamefont
			{Hollenberg}}, \ and\ \bibinfo {author} {\bibfnamefont {M.~Y.}\ \bibnamefont
			{Simmons}},\ }\href {\doibase 10.1126/science.1214319} {\bibfield  {journal}
		{\bibinfo  {journal} {Science}\ }\textbf {\bibinfo {volume} {335}},\ \bibinfo
		{pages} {64} (\bibinfo {year} {2012})}\BibitemShut {NoStop}%
	\bibitem [{\citenamefont {Bleszynski-Jayich}\ \emph {et~al.}(2009)\citenamefont
		{Bleszynski-Jayich}, \citenamefont {Shanks}, \citenamefont {Peaudecerf},
		\citenamefont {Ginossar}, \citenamefont {von Oppen}, \citenamefont
		{Glazman},\ and\ \citenamefont {Harris}}]{Bleszynski-Jayich272}%
	\BibitemOpen
	\bibfield  {author} {\bibinfo {author} {\bibfnamefont {A.~C.}\ \bibnamefont
			{Bleszynski-Jayich}}, \bibinfo {author} {\bibfnamefont {W.~E.}\ \bibnamefont
			{Shanks}}, \bibinfo {author} {\bibfnamefont {B.}~\bibnamefont {Peaudecerf}},
		\bibinfo {author} {\bibfnamefont {E.}~\bibnamefont {Ginossar}}, \bibinfo
		{author} {\bibfnamefont {F.}~\bibnamefont {von Oppen}}, \bibinfo {author}
		{\bibfnamefont {L.}~\bibnamefont {Glazman}}, \ and\ \bibinfo {author}
		{\bibfnamefont {J.~G.~E.}\ \bibnamefont {Harris}},\ }\href {\doibase
		10.1126/science.1178139} {\bibfield  {journal} {\bibinfo  {journal}
			{Science}\ }\textbf {\bibinfo {volume} {326}},\ \bibinfo {pages} {272}
		(\bibinfo {year} {2009})}\BibitemShut {NoStop}%
	\bibitem [{\citenamefont {Ghimire}\ \emph {et~al.}(2012)\citenamefont
		{Ghimire}, \citenamefont {DiChiara}, \citenamefont {Sistrunk}, \citenamefont
		{Ndabashimiye}, \citenamefont {Szafruga}, \citenamefont {Mohammad},
		\citenamefont {Agostini}, \citenamefont {DiMauro},\ and\ \citenamefont
		{Reis}}]{Ghimire2012}%
	\BibitemOpen
	\bibfield  {author} {\bibinfo {author} {\bibfnamefont {S.}~\bibnamefont
			{Ghimire}}, \bibinfo {author} {\bibfnamefont {A.~D.}\ \bibnamefont
			{DiChiara}}, \bibinfo {author} {\bibfnamefont {E.}~\bibnamefont {Sistrunk}},
		\bibinfo {author} {\bibfnamefont {G.}~\bibnamefont {Ndabashimiye}}, \bibinfo
		{author} {\bibfnamefont {U.~B.}\ \bibnamefont {Szafruga}}, \bibinfo {author}
		{\bibfnamefont {A.}~\bibnamefont {Mohammad}}, \bibinfo {author}
		{\bibfnamefont {P.}~\bibnamefont {Agostini}}, \bibinfo {author}
		{\bibfnamefont {L.~F.}\ \bibnamefont {DiMauro}}, \ and\ \bibinfo {author}
		{\bibfnamefont {D.~A.}\ \bibnamefont {Reis}},\ }\href {\doibase
		10.1103/PhysRevA.85.043836} {\bibfield  {journal} {\bibinfo  {journal} {Phys.
				Rev. A}\ }\textbf {\bibinfo {volume} {85}},\ \bibinfo {pages} {043836}
		(\bibinfo {year} {2012})}\BibitemShut {NoStop}%
	\bibitem [{\citenamefont {Ghimire}\ \emph
		{et~al.}(2011{\natexlab{a}})\citenamefont {Ghimire}, \citenamefont
		{DiChiara}, \citenamefont {Sistrunk}, \citenamefont {Agostini}, \citenamefont
		{DiMauro},\ and\ \citenamefont {Reis}}]{Ghimire2011a}%
	\BibitemOpen
	\bibfield  {author} {\bibinfo {author} {\bibfnamefont {S.}~\bibnamefont
			{Ghimire}}, \bibinfo {author} {\bibfnamefont {A.~D.}\ \bibnamefont
			{DiChiara}}, \bibinfo {author} {\bibfnamefont {E.}~\bibnamefont {Sistrunk}},
		\bibinfo {author} {\bibfnamefont {P.}~\bibnamefont {Agostini}}, \bibinfo
		{author} {\bibfnamefont {L.~F.}\ \bibnamefont {DiMauro}}, \ and\ \bibinfo
		{author} {\bibfnamefont {D.~A.}\ \bibnamefont {Reis}},\ }\href {\doibase
		10.1038/nphys1847} {\bibfield  {journal} {\bibinfo  {journal} {Nature
				Physics}\ }\textbf {\bibinfo {volume} {7}},\ \bibinfo {pages} {138} (\bibinfo
		{year} {2011}{\natexlab{a}})}\BibitemShut {NoStop}%
	\bibitem [{\citenamefont {Murakami}, \citenamefont {Eckstein},\ and\
		\citenamefont {Werner}(2018)}]{Murakami2018}%
	\BibitemOpen
	\bibfield  {author} {\bibinfo {author} {\bibfnamefont {Y.}~\bibnamefont
			{Murakami}}, \bibinfo {author} {\bibfnamefont {M.}~\bibnamefont {Eckstein}},
		\ and\ \bibinfo {author} {\bibfnamefont {P.}~\bibnamefont {Werner}},\ }\href
	{\doibase 10.1103/PhysRevLett.121.057405} {\bibfield  {journal} {\bibinfo
			{journal} {Phys. Rev. Lett.}\ }\textbf {\bibinfo {volume} {121}},\ \bibinfo
		{pages} {057405} (\bibinfo {year} {2018})}\BibitemShut {NoStop}%
	\bibitem [{\citenamefont {Bartels}\ \emph {et~al.}(2000)\citenamefont
		{Bartels}, \citenamefont {Backus}, \citenamefont {Zeek}, \citenamefont
		{Misogutl}, \citenamefont {Vdovin}, \citenamefont {Christov}, \citenamefont
		{Murnane},\ and\ \citenamefont {Kapteyn}}]{HHGcontrol}%
	\BibitemOpen
	\bibfield  {author} {\bibinfo {author} {\bibfnamefont {R.}~\bibnamefont
			{Bartels}}, \bibinfo {author} {\bibfnamefont {S.}~\bibnamefont {Backus}},
		\bibinfo {author} {\bibfnamefont {E.}~\bibnamefont {Zeek}}, \bibinfo {author}
		{\bibfnamefont {L.}~\bibnamefont {Misogutl}}, \bibinfo {author}
		{\bibfnamefont {G.}~\bibnamefont {Vdovin}}, \bibinfo {author} {\bibfnamefont
			{I.~P.}\ \bibnamefont {Christov}}, \bibinfo {author} {\bibfnamefont {M.~M.}\
			\bibnamefont {Murnane}}, \ and\ \bibinfo {author} {\bibfnamefont {H.~C.}\
			\bibnamefont {Kapteyn}},\ }\href {\doibase 10.1038/35018029} {\bibfield
		{journal} {\bibinfo  {journal} {Nature}\ }\textbf {\bibinfo {volume} {406}},\
		\bibinfo {pages} {164} (\bibinfo {year} {2000})}\BibitemShut {NoStop}%
	\bibitem [{\citenamefont {Rabitz}, \citenamefont {Hsieh},\ and\ \citenamefont
		{Rosenthal}(2004)}]{RabitzScience}%
	\BibitemOpen
	\bibfield  {author} {\bibinfo {author} {\bibfnamefont {H.~A.}\ \bibnamefont
			{Rabitz}}, \bibinfo {author} {\bibfnamefont {M.~M.}\ \bibnamefont {Hsieh}}, \
		and\ \bibinfo {author} {\bibfnamefont {C.~M.}\ \bibnamefont {Rosenthal}},\
	}\href@noop {} {\bibfield  {journal} {\bibinfo  {journal} {Science}\ }\textbf
		{\bibinfo {volume} {303}},\ \bibinfo {pages} {1998} (\bibinfo {year}
		{2004})}\BibitemShut {NoStop}%
	\bibitem [{\citenamefont {Werschnik}\ and\ \citenamefont
		{Gross}(2007)}]{Werschnik2007}%
	\BibitemOpen
	\bibfield  {author} {\bibinfo {author} {\bibfnamefont {J.}~\bibnamefont
			{Werschnik}}\ and\ \bibinfo {author} {\bibfnamefont {E.~K. U.}\ \bibnamefont
			{Gross}},\ }\href@noop {} {\bibfield  {journal} {\bibinfo  {journal} {J.
				Phys. B}\ }\textbf {\bibinfo {volume} {40}} (\bibinfo {year}
		{2007})}\BibitemShut {NoStop}%
	\bibitem [{\citenamefont {Serban}, \citenamefont {Werschnik},\ and\
		\citenamefont {Gross}(2005)}]{Serban2005}%
	\BibitemOpen
	\bibfield  {author} {\bibinfo {author} {\bibfnamefont {I.}~\bibnamefont
			{Serban}}, \bibinfo {author} {\bibfnamefont {J.}~\bibnamefont {Werschnik}}, \
		and\ \bibinfo {author} {\bibfnamefont {E.~K. U}\ \bibnamefont {Gross}},\ }\href
	{\doibase 10.1103/PhysRevA.71.053810} {\bibfield  {journal} {\bibinfo
			{journal} {Phys. Rev. A}\ }\textbf {\bibinfo {volume} {71}},\ \bibinfo
		{pages} {053810} (\bibinfo {year} {2005})}\BibitemShut {NoStop}%
	\bibitem [{\citenamefont {Palao}, \citenamefont {Kosloff},\ and\ \citenamefont
		{Koch}(2008)}]{Kosloffoptimalconstraints}%
	\BibitemOpen
	\bibfield  {author} {\bibinfo {author} {\bibfnamefont {J.~P.}\ \bibnamefont
			{Palao}}, \bibinfo {author} {\bibfnamefont {R.}~\bibnamefont {Kosloff}}, \
		and\ \bibinfo {author} {\bibfnamefont {C.~P.}\ \bibnamefont {Koch}},\ }\href
	{\doibase 10.1103/PhysRevA.77.063412} {\bibfield  {journal} {\bibinfo
			{journal} {Phys. Rev. A}\ }\textbf {\bibinfo {volume} {77}},\ \bibinfo
		{pages} {063412} (\bibinfo {year} {2008})}\BibitemShut {NoStop}%
	\bibitem [{\citenamefont {Doria}, \citenamefont {Calarco},\ and\ \citenamefont
		{Montangero}(2011)}]{PhysRevLett.106.190501}%
	\BibitemOpen
	\bibfield  {author} {\bibinfo {author} {\bibfnamefont {P.}~\bibnamefont
			{Doria}}, \bibinfo {author} {\bibfnamefont {T.}~\bibnamefont {Calarco}}, \
		and\ \bibinfo {author} {\bibfnamefont {S.}~\bibnamefont {Montangero}},\
	}\href {\doibase 10.1103/PhysRevLett.106.190501} {\bibfield  {journal}
		{\bibinfo  {journal} {Phys. Rev. Lett.}\ }\textbf {\bibinfo {volume} {106}},\
		\bibinfo {pages} {190501} (\bibinfo {year} {2011})}\BibitemShut {NoStop}%
	\bibitem [{\citenamefont {Kosloff}, \citenamefont {Hammerich},\ and\
		\citenamefont {Tannor}(1992)}]{Kosloff1992}%
	\BibitemOpen
	\bibfield  {author} {\bibinfo {author} {\bibfnamefont {R.}~\bibnamefont
			{Kosloff}}, \bibinfo {author} {\bibfnamefont {A.~D.}\ \bibnamefont
			{Hammerich}}, \ and\ \bibinfo {author} {\bibfnamefont {D.}~\bibnamefont
			{Tannor}},\ }\href {\doibase 10.1103/PhysRevLett.69.2172} {\bibfield
		{journal} {\bibinfo  {journal} {Phys. Rev. Lett.}\ }\textbf {\bibinfo
			{volume} {69}},\ \bibinfo {pages} {2172} (\bibinfo {year}
		{1992})}\BibitemShut {NoStop}%
	\bibitem [{\citenamefont {Bartana}, \citenamefont {Kosloff},\ and\
		\citenamefont {Tannor}(1993)}]{Koslofflocalcontrol}%
	\BibitemOpen
	\bibfield  {author} {\bibinfo {author} {\bibfnamefont {A.}~\bibnamefont
			{Bartana}}, \bibinfo {author} {\bibfnamefont {R.}~\bibnamefont {Kosloff}}, \
		and\ \bibinfo {author} {\bibfnamefont {D.~J.}\ \bibnamefont {Tannor}},\
	}\href {\doibase 10.1063/1.465797} {\bibfield  {journal} {\bibinfo  {journal}
			{J. Chem. Phys.}\ }\textbf {\bibinfo {volume} {99}},\ \bibinfo {pages} {196}
		(\bibinfo {year} {1993})}\BibitemShut {NoStop}%
	\bibitem [{\citenamefont {Zhang}\ \emph {et~al.}(2005)\citenamefont {Zhang},
		\citenamefont {Keys}, \citenamefont {Chen},\ and\ \citenamefont
		{Glotzer}}]{Zhang2005}%
	\BibitemOpen
	\bibfield  {author} {\bibinfo {author} {\bibnamefont {Zhang}}, \bibinfo
		{author} {\bibfnamefont {A.~S.}\ \bibnamefont {Keys}}, \bibinfo {author}
		{\bibfnamefont {T.}~\bibnamefont {Chen}}, \ and\ \bibinfo {author}
		{\bibfnamefont {S.~C.}\ \bibnamefont {Glotzer}},\ }\href {\doibase
		10.1021/la0513611} {\bibfield  {journal} {\bibinfo  {journal} {Langmuir}\
		}\textbf {\bibinfo {volume} {21}},\ \bibinfo {pages} {11547} (\bibinfo {year}
		{2005})}\BibitemShut {NoStop}%
	\bibitem [{\citenamefont {Whitesell}\ \emph {et~al.}(1994)\citenamefont
		{Whitesell}, \citenamefont {Davis}, \citenamefont {Wong},\ and\ \citenamefont
		{Chang}}]{doi:10.1021/ja00081a012}%
	\BibitemOpen
	\bibfield  {author} {\bibinfo {author} {\bibfnamefont {J.~K.}\ \bibnamefont
			{Whitesell}}, \bibinfo {author} {\bibfnamefont {R.~E.}\ \bibnamefont
			{Davis}}, \bibinfo {author} {\bibfnamefont {M.~S.}\ \bibnamefont {Wong}}, \
		and\ \bibinfo {author} {\bibfnamefont {N.~L.}\ \bibnamefont {Chang}},\
	}\href@noop {} {\bibfield  {journal} {\bibinfo  {journal} {J. Am. Chem.
				Soc.}\ }\textbf {\bibinfo {volume} {116}},\ \bibinfo {pages} {523} (\bibinfo
		{year} {1994})}\BibitemShut {NoStop}%
	\bibitem [{\citenamefont {Gust}, \citenamefont {Moore},\ and\ \citenamefont
		{Moore}(1993)}]{doi:10.1021/ar00028a010}%
	\BibitemOpen
	\bibfield  {author} {\bibinfo {author} {\bibfnamefont {D.}~\bibnamefont
			{Gust}}, \bibinfo {author} {\bibfnamefont {T.~A.}\ \bibnamefont {Moore}}, \
		and\ \bibinfo {author} {\bibfnamefont {A.~L.}\ \bibnamefont {Moore}},\
	}\href@noop {} {\bibfield  {journal} {\bibinfo  {journal} {Accs. Chem. Res.}\
		}\textbf {\bibinfo {volume} {26}},\ \bibinfo {pages} {198} (\bibinfo {year}
		{1993})}\BibitemShut {NoStop}%
	\bibitem [{\citenamefont {Rabong}\ \emph {et~al.}(2014)\citenamefont {Rabong},
		\citenamefont {Schuster}, \citenamefont {Liptaj}, \citenamefont {Pronayova},
		\citenamefont {Delchev}, \citenamefont {Jordis},\ and\ \citenamefont
		{Phopase}}]{C4RA01210K}%
	\BibitemOpen
	\bibfield  {author} {\bibinfo {author} {\bibfnamefont {C.}~\bibnamefont
			{Rabong}}, \bibinfo {author} {\bibfnamefont {C.}~\bibnamefont {Schuster}},
		\bibinfo {author} {\bibfnamefont {T.}~\bibnamefont {Liptaj}}, \bibinfo
		{author} {\bibfnamefont {N.}~\bibnamefont {Pronayova}}, \bibinfo {author}
		{\bibfnamefont {V.~B.}\ \bibnamefont {Delchev}}, \bibinfo {author}
		{\bibfnamefont {U.}~\bibnamefont {Jordis}}, \ and\ \bibinfo {author}
		{\bibfnamefont {J.}~\bibnamefont {Phopase}},\ }\href {\doibase
		10.1039/C4RA01210K} {\bibfield  {journal} {\bibinfo  {journal} {RSC Adv.}\
		}\textbf {\bibinfo {volume} {4}},\ \bibinfo {pages} {21351} (\bibinfo {year}
		{2014})}\BibitemShut {NoStop}%
	\bibitem [{\citenamefont {Della~Gaspera}\ \emph {et~al.}(2014)\citenamefont
		{Della~Gaspera}, \citenamefont {van Embden}, \citenamefont {Chesman},
		\citenamefont {Duffy},\ and\ \citenamefont
		{Jasieniak}}]{doi:10.1021/am506611j}%
	\BibitemOpen
	\bibfield  {author} {\bibinfo {author} {\bibfnamefont {E.}~\bibnamefont
			{Della~Gaspera}}, \bibinfo {author} {\bibfnamefont {J.}~\bibnamefont {van
				Embden}}, \bibinfo {author} {\bibfnamefont {A.~S.~R.}\ \bibnamefont
			{Chesman}}, \bibinfo {author} {\bibfnamefont {N.~W.}\ \bibnamefont {Duffy}},
		\ and\ \bibinfo {author} {\bibfnamefont {J.~J.}\ \bibnamefont {Jasieniak}},\
	}\href {\doibase 10.1021/am506611j} {\bibfield  {journal} {\bibinfo
			{journal} {ACS Appl. Mater. Interfaces}\ }\textbf {\bibinfo {volume} {6}},\
		\bibinfo {pages} {22519} (\bibinfo {year} {2014})}\BibitemShut {NoStop}%
	\bibitem [{\citenamefont {Nicoletti}\ \emph {et~al.}(2018)\citenamefont
		{Nicoletti}, \citenamefont {Fu}, \citenamefont {Mehio}, \citenamefont
		{Moore}, \citenamefont {Disa}, \citenamefont {Gu},\ and\ \citenamefont
		{Cavalleri}}]{Nicoletti2018}%
	\BibitemOpen
	\bibfield  {author} {\bibinfo {author} {\bibfnamefont {D.}~\bibnamefont
			{Nicoletti}}, \bibinfo {author} {\bibfnamefont {D.}~\bibnamefont {Fu}},
		\bibinfo {author} {\bibfnamefont {O.}~\bibnamefont {Mehio}}, \bibinfo
		{author} {\bibfnamefont {S.}~\bibnamefont {Moore}}, \bibinfo {author}
		{\bibfnamefont {A.~S.}\ \bibnamefont {Disa}}, \bibinfo {author}
		{\bibfnamefont {G.~D.}\ \bibnamefont {Gu}}, \ and\ \bibinfo {author}
		{\bibfnamefont {A.}~\bibnamefont {Cavalleri}},\ }\href {\doibase
		10.1103/PhysRevLett.121.267003} {\bibfield  {journal} {\bibinfo  {journal}
			{Phys. Rev. Lett.}\ }\textbf {\bibinfo {volume} {121}},\ \bibinfo {pages}
		{267003} (\bibinfo {year} {2018})}\BibitemShut {NoStop}%
	\bibitem [{\citenamefont {Schlawin}, \citenamefont {Cavalleri},\ and\
		\citenamefont {Jaksch}(2019)}]{Schlawin2019}%
	\BibitemOpen
	\bibfield  {author} {\bibinfo {author} {\bibfnamefont {F.}~\bibnamefont
			{Schlawin}}, \bibinfo {author} {\bibfnamefont {A.}~\bibnamefont {Cavalleri}},
		\ and\ \bibinfo {author} {\bibfnamefont {D.}~\bibnamefont {Jaksch}},\ }\href
	{\doibase 10.1103/PhysRevLett.122.133602} {\bibfield  {journal} {\bibinfo
			{journal} {Phys. Rev. Lett.}\ }\textbf {\bibinfo {volume} {122}},\ \bibinfo
		{pages} {133602} (\bibinfo {year} {2019})}\BibitemShut {NoStop}%
	\bibitem [{\citenamefont {Cavalleri}(2018)}]{Cavalleri}%
	\BibitemOpen
	\bibfield  {author} {\bibinfo {author} {\bibfnamefont {A.}~\bibnamefont
			{Cavalleri}},\ }\href@noop {} {\bibfield  {journal} {\bibinfo  {journal}
			{Contemp. Phys.}\ }\textbf {\bibinfo {volume} {59}},\ \bibinfo {pages} {31}
		(\bibinfo {year} {2018})}\BibitemShut {NoStop}%
	\bibitem [{\citenamefont {Rothman}, \citenamefont {Ho},\ and\ \citenamefont
		{Rabitz}(2005)}]{PhysRevA.72.023416}%
	\BibitemOpen
	\bibfield  {author} {\bibinfo {author} {\bibfnamefont {A.}~\bibnamefont
			{Rothman}}, \bibinfo {author} {\bibfnamefont {T.-S.}\ \bibnamefont {Ho}}, \
		and\ \bibinfo {author} {\bibfnamefont {H.}~\bibnamefont {Rabitz}},\ }\href
	{\doibase 10.1103/PhysRevA.72.023416} {\bibfield  {journal} {\bibinfo
			{journal} {Phys. Rev. A}\ }\textbf {\bibinfo {volume} {72}},\ \bibinfo
		{pages} {023416} (\bibinfo {year} {2005})}\BibitemShut {NoStop}%
	\bibitem [{\citenamefont {Magann}, \citenamefont {Ho},\ and\ \citenamefont
		{Rabitz}(2018)}]{PhysRevA.98.043429}%
	\BibitemOpen
	\bibfield  {author} {\bibinfo {author} {\bibfnamefont {A.}~\bibnamefont
			{Magann}}, \bibinfo {author} {\bibfnamefont {T.-S.}\ \bibnamefont {Ho}}, \
		and\ \bibinfo {author} {\bibfnamefont {H.}~\bibnamefont {Rabitz}},\ }\href
	{\doibase 10.1103/PhysRevA.98.043429} {\bibfield  {journal} {\bibinfo
			{journal} {Phys. Rev. A}\ }\textbf {\bibinfo {volume} {98}},\ \bibinfo
		{pages} {043429} (\bibinfo {year} {2018})}\BibitemShut {NoStop}%
	\bibitem [{\citenamefont {Caneva}, \citenamefont {Calarco},\ and\ \citenamefont
		{Montangero}(2011)}]{PhysRevA.84.022326}%
	\BibitemOpen
	\bibfield  {author} {\bibinfo {author} {\bibfnamefont {T.}~\bibnamefont
			{Caneva}}, \bibinfo {author} {\bibfnamefont {T.}~\bibnamefont {Calarco}}, \
		and\ \bibinfo {author} {\bibfnamefont {S.}~\bibnamefont {Montangero}},\
	}\href {\doibase 10.1103/PhysRevA.84.022326} {\bibfield  {journal} {\bibinfo
			{journal} {Phys. Rev. A}\ }\textbf {\bibinfo {volume} {84}},\ \bibinfo
		{pages} {022326} (\bibinfo {year} {2011})}\BibitemShut {NoStop}%
	\bibitem [{\citenamefont {Campos}\ \emph {et~al.}(2017)\citenamefont {Campos},
		\citenamefont {Bondar}, \citenamefont {Cabrera},\ and\ \citenamefont
		{Rabitz}}]{Campos2017}%
	\BibitemOpen
	\bibfield  {author} {\bibinfo {author} {\bibfnamefont {A.~G.}\ \bibnamefont
			{Campos}}, \bibinfo {author} {\bibfnamefont {D.~I.}\ \bibnamefont {Bondar}},
		\bibinfo {author} {\bibfnamefont {R.}~\bibnamefont {Cabrera}}, \ and\
		\bibinfo {author} {\bibfnamefont {H.~A.}\ \bibnamefont {Rabitz}},\ }\href
	{\doibase 10.1103/PhysRevLett.118.083201} {\bibfield  {journal} {\bibinfo
			{journal} {Phys. Rev. Lett.}\ }\textbf {\bibinfo {volume} {118}},\ \bibinfo
		{pages} {083201} (\bibinfo {year} {2017})}\BibitemShut {NoStop}%
	\bibitem [{\citenamefont {Zhu}\ and\ \citenamefont
		{Rabitz}(2003)}]{doi:10.1063/1.1582847}%
	\BibitemOpen
	\bibfield  {author} {\bibinfo {author} {\bibfnamefont {W.}~\bibnamefont
			{Zhu}}\ and\ \bibinfo {author} {\bibfnamefont {H.}~\bibnamefont {Rabitz}},\
	}\href {\doibase 10.1063/1.1582847} {\bibfield  {journal} {\bibinfo
			{journal} {J. Chem. Phys.}\ }\textbf {\bibinfo {volume} {119}},\ \bibinfo
		{pages} {3619} (\bibinfo {year} {2003})}\BibitemShut {NoStop}%
	\bibitem [{\citenamefont {Zhu}, \citenamefont {Smit},\ and\ \citenamefont
		{Rabitz}(1999)}]{doi:10.1063/1.477857}%
	\BibitemOpen
	\bibfield  {author} {\bibinfo {author} {\bibfnamefont {W.}~\bibnamefont
			{Zhu}}, \bibinfo {author} {\bibfnamefont {M.}~\bibnamefont {Smit}}, \ and\
		\bibinfo {author} {\bibfnamefont {H.}~\bibnamefont {Rabitz}},\ }\href
	{\doibase 10.1063/1.477857} {\bibfield  {journal} {\bibinfo  {journal} {J.
				Chem. Phys.}\ }\textbf {\bibinfo {volume} {110}},\ \bibinfo {pages} {1905}
		(\bibinfo {year} {1999})}\BibitemShut {NoStop}%
	\bibitem [{\citenamefont {McCaul}\ \emph {et~al.}(2019)\citenamefont {McCaul},
		\citenamefont {Orthodoxou}, \citenamefont {Jacobs}, \citenamefont {Booth},\
		and\ \citenamefont {Bondar}}]{companionlong}%
	\BibitemOpen
	\bibfield  {author} {\bibinfo {author} {\bibfnamefont {G.}~\bibnamefont
			{McCaul}}, \bibinfo {author} {\bibfnamefont {C.}~\bibnamefont {Orthodoxou}},
		\bibinfo {author} {\bibfnamefont {K.}~\bibnamefont {Jacobs}}, \bibinfo
		{author} {\bibfnamefont {G.~H.}\ \bibnamefont {Booth}}, \ and\ \bibinfo
		{author} {\bibfnamefont {D.~I.}\ \bibnamefont {Bondar}},\ }\href@noop {}
	{\bibfield  {journal} {\bibinfo  {journal} {Forthcoming}\ }\textbf {\bibinfo
			{volume} {0}},\ \bibinfo {pages} {0} (\bibinfo {year} {2019})}\BibitemShut
	{NoStop}%
	\bibitem [{\citenamefont {Tasaki}(1998)}]{Tasaki1998}%
	\BibitemOpen
	\bibfield  {author} {\bibinfo {author} {\bibfnamefont {H.}~\bibnamefont
			{Tasaki}},\ }\href {\doibase 10.1088/0953-8984/10/20/004} {\bibfield
		{journal} {\bibinfo  {journal} {J. Phys. Cond. Mat.}\ }\textbf {\bibinfo
			{volume} {10}},\ \bibinfo {pages} {4353} (\bibinfo {year}
		{1998})}\BibitemShut {NoStop}%
	\bibitem [{\citenamefont {Essler}(2005)}]{fabianessler2005}%
	\BibitemOpen
	\bibfield  {author} {\bibinfo {author} {\bibfnamefont {F.~H.~L.}\
			\bibnamefont {Essler}},\ }\href {https://www.xarg.org/ref/a/0521802628/}
	{\emph {\bibinfo {title} {The One-Dimensional Hubbard Model}}}\ (\bibinfo
	{publisher} {Cambridge University Press},\ \bibinfo {year}
	{2005})\BibitemShut {NoStop}%
	\bibitem [{\citenamefont {Folland}(2007)}]{geraldfolland2007}%
	\BibitemOpen
	\bibfield  {author} {\bibinfo {author} {\bibfnamefont {G.~B.}\ \bibnamefont
			{Folland}},\ }\href {https://www.xarg.org/ref/a/0471317160/} {\emph {\bibinfo
			{title} {Real Analysis: Modern Techniques and Their Applications}}}\
	(\bibinfo  {publisher} {Wiley},\ \bibinfo {year} {2007})\BibitemShut
	{NoStop}%
	\bibitem [{\citenamefont {Nagle}(2011)}]{kentnagle2011}%
	\BibitemOpen
	\bibfield  {author} {\bibinfo {author} {\bibfnamefont {R.~K.}\ \bibnamefont
			{Nagle}},\ }\href {https://www.xarg.org/ref/a/0321747747/} {\emph {\bibinfo
			{title} {Fundamentals of Differential Equations and Boundary Value Problems
				(6th Edition) (Featured Titles for Differential Equations)}}}\ (\bibinfo
	{publisher} {Pearson},\ \bibinfo {year} {2011})\BibitemShut {NoStop}%
	\bibitem [{\citenamefont {Jha}\ \emph {et~al.}(2009)\citenamefont {Jha},
		\citenamefont {Beltrani}, \citenamefont {Rosenthal},\ and\ \citenamefont
		{Rabitz}}]{Rabitzmultiple}%
	\BibitemOpen
	\bibfield  {author} {\bibinfo {author} {\bibfnamefont {A.}~\bibnamefont
			{Jha}}, \bibinfo {author} {\bibfnamefont {V.}~\bibnamefont {Beltrani}},
		\bibinfo {author} {\bibfnamefont {C.}~\bibnamefont {Rosenthal}}, \ and\
		\bibinfo {author} {\bibfnamefont {H.}~\bibnamefont {Rabitz}},\ }\href
	{\doibase 10.1021/jp811485j} {\bibfield  {journal} {\bibinfo  {journal} {J.
				Phys. Chem. A}\ }\textbf {\bibinfo {volume} {113}},\ \bibinfo {pages} {7667}
		(\bibinfo {year} {2009})}\BibitemShut {NoStop}%
	\bibitem [{\citenamefont {Hirschorn}\ and\ \citenamefont
		{Davis}(1987)}]{doi:10.1137/0325030}%
	\BibitemOpen
	\bibfield  {author} {\bibinfo {author} {\bibfnamefont {R.}~\bibnamefont
			{Hirschorn}}\ and\ \bibinfo {author} {\bibfnamefont {J.}~\bibnamefont
			{Davis}},\ }\href@noop {} {\bibfield  {journal} {\bibinfo  {journal} {SIAM J.
				Comput.}\ }\textbf {\bibinfo {volume} {25}},\ \bibinfo {pages} {547}
		(\bibinfo {year} {1987})}\BibitemShut {NoStop}%
	\bibitem [{\citenamefont {Haldane}(1980)}]{PhysRevLett.45.1358}%
	\BibitemOpen
	\bibfield  {author} {\bibinfo {author} {\bibfnamefont {F.~D.~M.}\
			\bibnamefont {Haldane}},\ }\href {\doibase 10.1103/PhysRevLett.45.1358}
	{\bibfield  {journal} {\bibinfo  {journal} {Phys. Rev. Lett.}\ }\textbf
		{\bibinfo {volume} {45}},\ \bibinfo {pages} {1358} (\bibinfo {year}
		{1980})}\BibitemShut {NoStop}%
	\bibitem [{\citenamefont {Jeckelmann}, \citenamefont {Gebhard},\ and\
		\citenamefont {Essler}(2000)}]{PhysRevLett.85.3910}%
	\BibitemOpen
	\bibfield  {author} {\bibinfo {author} {\bibfnamefont {E.}~\bibnamefont
			{Jeckelmann}}, \bibinfo {author} {\bibfnamefont {F.}~\bibnamefont {Gebhard}},
		\ and\ \bibinfo {author} {\bibfnamefont {F.~H.~L.}\ \bibnamefont {Essler}},\
	}\href {\doibase 10.1103/PhysRevLett.85.3910} {\bibfield  {journal} {\bibinfo
			{journal} {Phys. Rev. Lett.}\ }\textbf {\bibinfo {volume} {85}},\ \bibinfo
		{pages} {3910} (\bibinfo {year} {2000})}\BibitemShut {NoStop}%
	\bibitem [{\citenamefont {Silva}\ \emph {et~al.}(2018)\citenamefont {Silva},
		\citenamefont {Blinov}, \citenamefont {Rubtsov}, \citenamefont {Smirnova},\
		and\ \citenamefont {Ivanov}}]{Silva2018}%
	\BibitemOpen
	\bibfield  {author} {\bibinfo {author} {\bibfnamefont {R.~E.}\ \bibnamefont
			{Silva}}, \bibinfo {author} {\bibfnamefont {I.~V.}\ \bibnamefont {Blinov}},
		\bibinfo {author} {\bibfnamefont {A.~N.}\ \bibnamefont {Rubtsov}}, \bibinfo
		{author} {\bibfnamefont {O.}~\bibnamefont {Smirnova}}, \ and\ \bibinfo
		{author} {\bibfnamefont {M.}~\bibnamefont {Ivanov}},\ }\href@noop {}
	{\bibfield  {journal} {\bibinfo  {journal} {Nature Photonics}\ }\textbf
		{\bibinfo {volume} {12}},\ \bibinfo {pages} {266} (\bibinfo {year}
		{2018})}\BibitemShut {NoStop}%
	\bibitem [{\citenamefont {Hohenleutner}\ \emph {et~al.}(2015)\citenamefont
		{Hohenleutner}, \citenamefont {Langer}, \citenamefont {Schubert},
		\citenamefont {Knorr}, \citenamefont {Huttner}, \citenamefont {Koch},
		\citenamefont {Kira},\ and\ \citenamefont {Huber}}]{Hohenleutner2015}%
	\BibitemOpen
	\bibfield  {author} {\bibinfo {author} {\bibfnamefont {M.}~\bibnamefont
			{Hohenleutner}}, \bibinfo {author} {\bibfnamefont {F.}~\bibnamefont
			{Langer}}, \bibinfo {author} {\bibfnamefont {O.}~\bibnamefont {Schubert}},
		\bibinfo {author} {\bibfnamefont {M.}~\bibnamefont {Knorr}}, \bibinfo
		{author} {\bibfnamefont {U.}~\bibnamefont {Huttner}}, \bibinfo {author}
		{\bibfnamefont {S.~W.}\ \bibnamefont {Koch}}, \bibinfo {author}
		{\bibfnamefont {M.}~\bibnamefont {Kira}}, \ and\ \bibinfo {author}
		{\bibfnamefont {R.}~\bibnamefont {Huber}},\ }\href@noop {} {\bibfield
		{journal} {\bibinfo  {journal} {Nature}\ }\textbf {\bibinfo {volume} {523}},\
		\bibinfo {pages} {572 EP } (\bibinfo {year} {2015})}\BibitemShut {NoStop}%
	\bibitem [{\citenamefont {Ghimire}\ \emph
		{et~al.}(2011{\natexlab{b}})\citenamefont {Ghimire}, \citenamefont
		{DiChiara}, \citenamefont {Sistrunk}, \citenamefont {Szafruga}, \citenamefont
		{Agostini}, \citenamefont {DiMauro},\ and\ \citenamefont
		{Reis}}]{Ghimire2011}%
	\BibitemOpen
	\bibfield  {author} {\bibinfo {author} {\bibfnamefont {S.}~\bibnamefont
			{Ghimire}}, \bibinfo {author} {\bibfnamefont {A.~D.}\ \bibnamefont
			{DiChiara}}, \bibinfo {author} {\bibfnamefont {E.}~\bibnamefont {Sistrunk}},
		\bibinfo {author} {\bibfnamefont {U.~B.}\ \bibnamefont {Szafruga}}, \bibinfo
		{author} {\bibfnamefont {P.}~\bibnamefont {Agostini}}, \bibinfo {author}
		{\bibfnamefont {L.~F.}\ \bibnamefont {DiMauro}}, \ and\ \bibinfo {author}
		{\bibfnamefont {D.~A.}\ \bibnamefont {Reis}},\ }\href {\doibase
		10.1103/PhysRevLett.107.167407} {\bibfield  {journal} {\bibinfo  {journal}
			{Phys. Rev. Lett.}\ }\textbf {\bibinfo {volume} {107}},\ \bibinfo {pages}
		{167407} (\bibinfo {year} {2011}{\natexlab{b}})}\BibitemShut {NoStop}%
	\bibitem [{\citenamefont {P{\'{e}}pin}\ \emph {et~al.}(2004)\citenamefont
		{P{\'{e}}pin}, \citenamefont {Niikura}, \citenamefont {Corkum}, \citenamefont
		{Villeneuve}, \citenamefont {Kieffer}, \citenamefont {Levesque},
		\citenamefont {Itatani},\ and\ \citenamefont {Zeidler}}]{Pepin2004}%
	\BibitemOpen
	\bibfield  {author} {\bibinfo {author} {\bibfnamefont {H.}~\bibnamefont
			{P{\'{e}}pin}}, \bibinfo {author} {\bibfnamefont {H.}~\bibnamefont
			{Niikura}}, \bibinfo {author} {\bibfnamefont {P.~B.}\ \bibnamefont {Corkum}},
		\bibinfo {author} {\bibfnamefont {D.~M.}\ \bibnamefont {Villeneuve}},
		\bibinfo {author} {\bibfnamefont {J.~C.}\ \bibnamefont {Kieffer}}, \bibinfo
		{author} {\bibfnamefont {J.}~\bibnamefont {Levesque}}, \bibinfo {author}
		{\bibfnamefont {J.}~\bibnamefont {Itatani}}, \ and\ \bibinfo {author}
		{\bibfnamefont {D.}~\bibnamefont {Zeidler}},\ }\href {\doibase
		10.1038/nature03183} {\bibfield  {journal} {\bibinfo  {journal} {Nature}\
		}\textbf {\bibinfo {volume} {432}},\ \bibinfo {pages} {867} (\bibinfo {year}
		{2004})}\BibitemShut {NoStop}%
	\bibitem [{\citenamefont {Krausz}\ and\ \citenamefont
		{Ivanov}(2009)}]{RevModPhys.81.163}%
	\BibitemOpen
	\bibfield  {author} {\bibinfo {author} {\bibfnamefont {F.}~\bibnamefont
			{Krausz}}\ and\ \bibinfo {author} {\bibfnamefont {M.}~\bibnamefont
			{Ivanov}},\ }\href {\doibase 10.1103/RevModPhys.81.163} {\bibfield  {journal}
		{\bibinfo  {journal} {Rev. Mod. Phys.}\ }\textbf {\bibinfo {volume} {81}},\
		\bibinfo {pages} {163} (\bibinfo {year} {2009})}\BibitemShut {NoStop}%
	\bibitem [{\citenamefont {Neufeld}\ \emph {et~al.}(2019)\citenamefont
		{Neufeld}, \citenamefont {Ayuso}, \citenamefont {Decleva}, \citenamefont
		{Ivanov}, \citenamefont {Smirnova},\ and\ \citenamefont
		{Cohen}}]{PhysRevX.9.031002}%
	\BibitemOpen
	\bibfield  {author} {\bibinfo {author} {\bibfnamefont {O.}~\bibnamefont
			{Neufeld}}, \bibinfo {author} {\bibfnamefont {D.}~\bibnamefont {Ayuso}},
		\bibinfo {author} {\bibfnamefont {P.}~\bibnamefont {Decleva}}, \bibinfo
		{author} {\bibfnamefont {M.~Y.}\ \bibnamefont {Ivanov}}, \bibinfo {author}
		{\bibfnamefont {O.}~\bibnamefont {Smirnova}}, \ and\ \bibinfo {author}
		{\bibfnamefont {O.}~\bibnamefont {Cohen}},\ }\href {\doibase
		10.1103/PhysRevX.9.031002} {\bibfield  {journal} {\bibinfo  {journal} {Phys.
				Rev. X}\ }\textbf {\bibinfo {volume} {9}},\ \bibinfo {pages} {031002}
		(\bibinfo {year} {2019})}\BibitemShut {NoStop}%
	\bibitem [{\citenamefont {Lieb}\ and\ \citenamefont
		{Wu}(1968)}]{PhysRevLett.20.1445}%
	\BibitemOpen
	\bibfield  {author} {\bibinfo {author} {\bibfnamefont {E.~H.}\ \bibnamefont
			{Lieb}}\ and\ \bibinfo {author} {\bibfnamefont {F.~Y.}\ \bibnamefont {Wu}},\
	}\href {\doibase 10.1103/PhysRevLett.20.1445} {\bibfield  {journal} {\bibinfo
			{journal} {Phys. Rev. Lett.}\ }\textbf {\bibinfo {volume} {20}},\ \bibinfo
		{pages} {1445} (\bibinfo {year} {1968})}\BibitemShut {NoStop}%
	\bibitem [{\citenamefont {Gebhard}(2010)}]{floriangebhard2010}%
	\BibitemOpen
	\bibfield  {author} {\bibinfo {author} {\bibfnamefont {F.}~\bibnamefont
			{Gebhard}},\ }\href {https://www.xarg.org/ref/a/3642082637/} {\emph {\bibinfo
			{title} {The Mott Metal-Insulator Transition: Models and Methods (Springer
				Tracts in Modern Physics)}}}\ (\bibinfo  {publisher} {Springer},\ \bibinfo
	{year} {2010})\BibitemShut {NoStop}%
	\bibitem [{\citenamefont {Oka}(2012)}]{PhysRevB.86.075148}%
	\BibitemOpen
	\bibfield  {author} {\bibinfo {author} {\bibfnamefont {T.}~\bibnamefont
			{Oka}},\ }\href {\doibase 10.1103/PhysRevB.86.075148} {\bibfield  {journal}
		{\bibinfo  {journal} {Phys. Rev. B}\ }\textbf {\bibinfo {volume} {86}},\
		\bibinfo {pages} {075148} (\bibinfo {year} {2012})}\BibitemShut {NoStop}%
	\bibitem [{\citenamefont {Leigh}\ and\ \citenamefont
		{Phillips}(2009)}]{PhysRevB.79.245120}%
	\BibitemOpen
	\bibfield  {author} {\bibinfo {author} {\bibfnamefont {R.~G.}\ \bibnamefont
			{Leigh}}\ and\ \bibinfo {author} {\bibfnamefont {P.}~\bibnamefont
			{Phillips}},\ }\href {\doibase 10.1103/PhysRevB.79.245120} {\bibfield
		{journal} {\bibinfo  {journal} {Phys. Rev. B}\ }\textbf {\bibinfo {volume}
			{79}},\ \bibinfo {pages} {245120} (\bibinfo {year} {2009})}\BibitemShut
	{NoStop}%
	\bibitem [{\citenamefont {Lieb}\ and\ \citenamefont {Wu}(2003)}]{LIEB20031}%
	\BibitemOpen
	\bibfield  {author} {\bibinfo {author} {\bibfnamefont {E.~H.}\ \bibnamefont
			{Lieb}}\ and\ \bibinfo {author} {\bibfnamefont {F.}~\bibnamefont {Wu}},\
	}\href {\doibase https://doi.org/10.1016/S0378-4371(02)01785-5} {\bibfield
		{journal} {\bibinfo  {journal} {Physica A}\ }\textbf {\bibinfo {volume}
			{321}},\ \bibinfo {pages} {1 } (\bibinfo {year} {2003})}\BibitemShut
	{NoStop}%
	\bibitem [{\citenamefont {Stafford}\ and\ \citenamefont
		{Millis}(1993)}]{PhysRevB.48.1409}%
	\BibitemOpen
	\bibfield  {author} {\bibinfo {author} {\bibfnamefont {C.~A.}\ \bibnamefont
			{Stafford}}\ and\ \bibinfo {author} {\bibfnamefont {A.~J.}\ \bibnamefont
			{Millis}},\ }\href {\doibase 10.1103/PhysRevB.48.1409} {\bibfield  {journal}
		{\bibinfo  {journal} {Phys. Rev. B}\ }\textbf {\bibinfo {volume} {48}},\
		\bibinfo {pages} {1409} (\bibinfo {year} {1993})}\BibitemShut {NoStop}%
	\bibitem [{\citenamefont {Schubert}\ \emph {et~al.}(2014)\citenamefont
		{Schubert}, \citenamefont {Hohenleutner}, \citenamefont {Langer},
		\citenamefont {Urbanek}, \citenamefont {Lange}, \citenamefont {Huttner},
		\citenamefont {Golde}, \citenamefont {Meier}, \citenamefont {Kira},
		\citenamefont {Koch},\ and\ \citenamefont {Huber}}]{Schubert2014}%
	\BibitemOpen
	\bibfield  {author} {\bibinfo {author} {\bibfnamefont {O.}~\bibnamefont
			{Schubert}}, \bibinfo {author} {\bibfnamefont {M.}~\bibnamefont
			{Hohenleutner}}, \bibinfo {author} {\bibfnamefont {F.}~\bibnamefont
			{Langer}}, \bibinfo {author} {\bibfnamefont {B.}~\bibnamefont {Urbanek}},
		\bibinfo {author} {\bibfnamefont {C.}~\bibnamefont {Lange}}, \bibinfo
		{author} {\bibfnamefont {U.}~\bibnamefont {Huttner}}, \bibinfo {author}
		{\bibfnamefont {D.}~\bibnamefont {Golde}}, \bibinfo {author} {\bibfnamefont
			{T.}~\bibnamefont {Meier}}, \bibinfo {author} {\bibfnamefont
			{M.}~\bibnamefont {Kira}}, \bibinfo {author} {\bibfnamefont {S.~W.}\
			\bibnamefont {Koch}}, \ and\ \bibinfo {author} {\bibfnamefont
			{R.}~\bibnamefont {Huber}},\ }\href {\doibase 10.1038/nphoton.2013.349}
	{\bibfield  {journal} {\bibinfo  {journal} {Nature Photonics}\ }\textbf
		{\bibinfo {volume} {8}},\ \bibinfo {pages} {119} (\bibinfo {year}
		{2014})}\BibitemShut {NoStop}%
	\bibitem [{\citenamefont {Hawkins}\ and\ \citenamefont
		{Ivanov}(2013)}]{Hawkins2013}%
	\BibitemOpen
	\bibfield  {author} {\bibinfo {author} {\bibfnamefont {P.~G.}\ \bibnamefont
			{Hawkins}}\ and\ \bibinfo {author} {\bibfnamefont {M.~Y.}\ \bibnamefont
			{Ivanov}},\ }\href {\doibase 10.1103/PhysRevA.87.063842} {\bibfield
		{journal} {\bibinfo  {journal} {Phys. Rev. A}\ }\textbf {\bibinfo {volume}
			{87}},\ \bibinfo {pages} {063842} (\bibinfo {year} {2013})}\BibitemShut
	{NoStop}%
		\bibitem [{\citenamefont {Liu}\ \emph {et~al.}(2018)\citenamefont {Liu},
  \citenamefont {Guo}, \citenamefont {Vampa}, \citenamefont {Zhang},
  \citenamefont {Sarmiento}, \citenamefont {Xiao}, \citenamefont {Bucksbaum},
  \citenamefont {Vu{\v{c}}kovi{\'{c}}}, \citenamefont {Fan},\ and\
  \citenamefont {Reis}}]{HHGenhance1}%
  \BibitemOpen
  \bibfield  {author} {\bibinfo {author} {\bibfnamefont {H.}~\bibnamefont
  {Liu}}, \bibinfo {author} {\bibfnamefont {C.}~\bibnamefont {Guo}}, \bibinfo
  {author} {\bibfnamefont {G.}~\bibnamefont {Vampa}}, \bibinfo {author}
  {\bibfnamefont {J.~L.}\ \bibnamefont {Zhang}}, \bibinfo {author}
  {\bibfnamefont {T.}~\bibnamefont {Sarmiento}}, \bibinfo {author}
  {\bibfnamefont {M.}~\bibnamefont {Xiao}}, \bibinfo {author} {\bibfnamefont
  {P.~H.}\ \bibnamefont {Bucksbaum}}, \bibinfo {author} {\bibfnamefont
  {J.}~\bibnamefont {Vu{\v{c}}kovi{\'{c}}}}, \bibinfo {author} {\bibfnamefont
  {S.}~\bibnamefont {Fan}}, \ and\ \bibinfo {author} {\bibfnamefont {D.~A.}\
  \bibnamefont {Reis}},\ }\href {\doibase 10.1038/s41567-018-0233-6} {\bibfield
   {journal} {\bibinfo  {journal} {Nature Physics}\ }\textbf {\bibinfo {volume}
  {14}},\ \bibinfo {pages} {1006} (\bibinfo {year} {2018})},\ \Eprint
  {http://arxiv.org/abs/1710.04244} {arXiv:1710.04244} \BibitemShut {NoStop}%
\bibitem [{\citenamefont {Jin}\ \emph {et~al.}(2014)\citenamefont {Jin},
  \citenamefont {Wang}, \citenamefont {Wei}, \citenamefont {Le},\ and\
  \citenamefont {Lin}}]{HHGenhance2}%
  \BibitemOpen
  \bibfield  {author} {\bibinfo {author} {\bibfnamefont {C.}~\bibnamefont
  {Jin}}, \bibinfo {author} {\bibfnamefont {G.}~\bibnamefont {Wang}}, \bibinfo
  {author} {\bibfnamefont {H.}~\bibnamefont {Wei}}, \bibinfo {author}
  {\bibfnamefont {A.~T.}\ \bibnamefont {Le}}, \ and\ \bibinfo {author}
  {\bibfnamefont {C.~D.}\ \bibnamefont {Lin}},\ }\href {\doibase
  10.1038/ncomms5003} {\bibfield  {journal} {\bibinfo  {journal} {Nature
  Communications}\ }\textbf {\bibinfo {volume} {5}},\ \bibinfo {pages} {4003}
  (\bibinfo {year} {2014})}\BibitemShut {NoStop}%
\bibitem [{\citenamefont {Kroh}\ \emph {et~al.}(2018)\citenamefont {Kroh},
  \citenamefont {Jin}, \citenamefont {Krogen}, \citenamefont {Keathley},
  \citenamefont {Calendron}, \citenamefont {Siqueira}, \citenamefont {Liang},
  \citenamefont {Falc{\~{a}}o-Filho}, \citenamefont {Lin}, \citenamefont
  {K{\"{a}}rtner},\ and\ \citenamefont {Hong}}]{HHGenhance3}%
  \BibitemOpen
  \bibfield  {author} {\bibinfo {author} {\bibfnamefont {T.}~\bibnamefont
  {Kroh}}, \bibinfo {author} {\bibfnamefont {C.}~\bibnamefont {Jin}}, \bibinfo
  {author} {\bibfnamefont {P.}~\bibnamefont {Krogen}}, \bibinfo {author}
  {\bibfnamefont {P.~D.}\ \bibnamefont {Keathley}}, \bibinfo {author}
  {\bibfnamefont {A.-L.}\ \bibnamefont {Calendron}}, \bibinfo {author}
  {\bibfnamefont {J.~P.}\ \bibnamefont {Siqueira}}, \bibinfo {author}
  {\bibfnamefont {H.}~\bibnamefont {Liang}}, \bibinfo {author} {\bibfnamefont
  {E.~L.}\ \bibnamefont {Falc{\~{a}}o-Filho}}, \bibinfo {author} {\bibfnamefont
  {C.~D.}\ \bibnamefont {Lin}}, \bibinfo {author} {\bibfnamefont {F.~X.}\
  \bibnamefont {K{\"{a}}rtner}}, \ and\ \bibinfo {author} {\bibfnamefont
  {K.-H.}\ \bibnamefont {Hong}},\ }\href {\doibase 10.1364/oe.26.016955}
  {\bibfield  {journal} {\bibinfo  {journal} {Optics Express}\ }\textbf
  {\bibinfo {volume} {26}},\ \bibinfo {pages} {16955} (\bibinfo {year}
  {2018})}\BibitemShut {NoStop}%
	\bibitem [{\citenamefont {Sommer}\ \emph {et~al.}(2016)\citenamefont {Sommer},
		\citenamefont {Bothschafter}, \citenamefont {Sato}, \citenamefont {Jakubeit},
		\citenamefont {Latka}, \citenamefont {Razskazovskaya}, \citenamefont
		{Fattahi}, \citenamefont {Jobst}, \citenamefont {Schweinberger},
		\citenamefont {Shirvanyan}, \citenamefont {Yakovlev}, \citenamefont
		{Kienberger}, \citenamefont {Yabana}, \citenamefont {Karpowicz},
		\citenamefont {Schultze},\ and\ \citenamefont {Krausz}}]{Sommer2016}%
	\BibitemOpen
	\bibfield  {author} {\bibinfo {author} {\bibfnamefont {A.}~\bibnamefont
			{Sommer}}, \bibinfo {author} {\bibfnamefont {E.~M.}\ \bibnamefont
			{Bothschafter}}, \bibinfo {author} {\bibfnamefont {S.~A.}\ \bibnamefont
			{Sato}}, \bibinfo {author} {\bibfnamefont {C.}~\bibnamefont {Jakubeit}},
		\bibinfo {author} {\bibfnamefont {T.}~\bibnamefont {Latka}}, \bibinfo
		{author} {\bibfnamefont {O.}~\bibnamefont {Razskazovskaya}}, \bibinfo
		{author} {\bibfnamefont {H.}~\bibnamefont {Fattahi}}, \bibinfo {author}
		{\bibfnamefont {M.}~\bibnamefont {Jobst}}, \bibinfo {author} {\bibfnamefont
			{W.}~\bibnamefont {Schweinberger}}, \bibinfo {author} {\bibfnamefont
			{V.}~\bibnamefont {Shirvanyan}}, \bibinfo {author} {\bibfnamefont {V.~S.}\
			\bibnamefont {Yakovlev}}, \bibinfo {author} {\bibfnamefont {R.}~\bibnamefont
			{Kienberger}}, \bibinfo {author} {\bibfnamefont {K.}~\bibnamefont {Yabana}},
		\bibinfo {author} {\bibfnamefont {N.}~\bibnamefont {Karpowicz}}, \bibinfo
		{author} {\bibfnamefont {M.}~\bibnamefont {Schultze}}, \ and\ \bibinfo
		{author} {\bibfnamefont {F.}~\bibnamefont {Krausz}},\ }\href {\doibase
		10.1038/nature17650} {\bibfield  {journal} {\bibinfo  {journal} {Nature}\
		}\textbf {\bibinfo {volume} {534}},\ \bibinfo {pages} {86} (\bibinfo {year}
		{2016})}\BibitemShut {NoStop}%
\end{thebibliography}
\end{document}